\documentclass[iop]{emulateapj}
\usepackage{rotating}
\usepackage{mathtools}
\usepackage{graphicx}
\usepackage{multirow}

\slugcomment{}

\shorttitle{}
\shortauthors{Pace \& Salim}

\begin{document}

\title{SUPPRESSION OF STAR FORMATION IN THE HOSTS OF LOW-EXCITATION RADIO GALAXIES }

\author{Cameron Pace\altaffilmark{1} \& Samir Salim }
\affil{Indiana University, Dept. of Astronomy, Swain Hall West 319, Bloomington, IN, USA 47405-7105}

\altaffiltext{1}{Current Address: Southern Utah University, Dept. of Physical Science, 351 W. University Blvd., Cedar City, UT, 84720}
\email{cameronpace@suu.edu, salims@indiana.edu}

\begin{abstract}
The feedback from radio-loud active galactic nuclei (R-AGN) may help maintain low star formation (SF) rates in their early-type hosts, but the observational evidence for this mechanism has been inconclusive. We study systematic differences of aggregate spectral energy distributions (SEDs) of various subsets of $\sim$4000 low-redshift R-AGN from Best \& Heckman (2012) with respect to (currently) inactive control samples selected to have matching redshift, stellar mass, population age, axis ratio, and environment. Aggregate SEDs, ranging from the ultraviolet (UV) through mid-infrared (mid-IR, 22 $\mu$m), were constructed using a Bayesian method that eliminates biases from non-detections in {\it GALEX} and {\it WISE}. We study rare high-excitation sources separately from low-excitation ones, which we split by environment and host properties.  We find that both the UV and mid-IR emission of non-cluster R-AGNs (80\% of sample) are suppressed by $\sim$0.2 dex relative to that of the control group, especially for moderately massive galaxies (log $M_* \lesssim$ 11). The difference disappears for high-mass R-AGN and for R-AGN in clusters, where other, non-AGN quenching/maintenance mechanisms may dominate, or where the suppression of SF due to AGN may persist between active phases of the central engine, perhaps because of the presence of a hot gaseous halo storing AGN energy.  High-excitation (high accretion rate) sources, which make up 2\% of the R-AGN sample, also show no evidence of SF suppression (their UV is the same as in controls), but they exhibit a strong mid-IR excess due to AGN dust heating. 
\end{abstract}


\section{Introduction}
\label{sec:intro}

Active galactic nuclei (AGN) are powered by the accretion of matter onto a supermassive black hole (SMBH). 
Radiatively efficient AGN, which includes optically and X-ray-selected AGN as well as luminous radio-loud AGN (R-AGN), are powered by the efficient accretion of matter from the accretion disk onto the SMBH. 
These  AGN may be triggered by mergers (Heckman et al. 1986, Barnes \& Hernquist 1996, Satyapal et al. 2014), although the lack of an environment density-AGN luminosity relation suggests that this may not be the case \citep{2014MNRAS.439..861K}.
While some R-AGN are found in late-type galaxies \citep{2015MNRAS.454.1595K,2015MNRAS.454.1556S}, the majority of R-AGN in the current universe are found in passively evolving galaxies \citep{2012MNRAS.421.1569B} and are powered by the radiatively inefficient accretion of hot gas via either Bondi accretion or the advective cooling of gas from the halo or from evolved stars \citep{1994ApJ...428L..13N, 2006MNRAS.372...21A, 2007ApJ...662..166H, 2008ARA&A..46..475H}.

Both modes of AGN may affect their host galaxies via feedback: radiatively efficient AGN via powerful winds, and radiatively inefficient AGN via their extended radio jets.
Although AGN-driven feedback could in principle boost a galaxy's star formation rate (SFR) by inducing the collapse of cold gas clouds \citep{2009ApJ...700..262S, 2014ApJ...796..113D}, its \textit{negative} effect on star formation (SF) is more commonly discussed in the literature. 
This negative feedback is required for simulations and semianalytic models to match the observed properties of massive ellipticals (e.g. Granato et al. 2004; Ciotti et al. 2010; Cattaneo et al. 2009) as well as prevent the overproduction of massive galaxies \citep{2006MNRAS.365...11C}. 
Many clusters whose brightest cluster galaxies (BCGs) host radio jets are also observed to have X-ray bubbles (Dunn \& Fabian 2006; 2008), which could indicate that heating by AGN balances cooling flows in these clusters \citep{1994ARA&A..32..277F}. This balance may depend on the timescales and instabilities of the cooling gas \citep{2015ApJ...808L..30V}.
Although this negative feedback may be observed in cluster environments, it is uncertain whether it takes place in individual galaxies, partly due to the greater difficulty in detecting the fainter X-ray emission from their lower-density halo gas \citep{2012ARA&A..50..455F}.

In this paper, we examine the effects of R-AGN activity on the stellar populations of their host galaxies using a large sample of R-AGN in the local universe, identified by matching radio surveys to optical, ultraviolet (UV) and mid-infrared (mid-IR) surveys. 
We are most interested in the effects of radio jets on recent star formation, which can be traced in the UV.
Mid-infrared  emission traces star formation, but it may also arise from old stellar populations or nuclear activity.
To disentangle the effects of these emission sources we compare the UV and mid-IR emission of a sample of R-AGN, drawn from the \citet{2012MNRAS.421.1569B} catalog, to that of a carefully selected sample of control galaxies.

Since R-AGN bubbles are observed in clusters but may not be present in field galaxies, we treat R-AGN in different environments (field galaxies, cluster members, BCGs) separately. 
Many studies find that R-AGN tend to be found in denser environments (e.g. Best 2004; Kauffmann et al. 2008; Reviglio \& Helfand 2006; Pace \& Salim 2014), but this general statement hides important details.  
The incidence rate of R-AGN is higher in BCGs than galaxies in the field \citep{1990AJ.....99...14B, 2005MNRAS.362...25B, 2012A&A...544A..18V}, although the incidence rate in cluster members is comparable to that found in the field \citep{2014ApJ...785...66P}. This suggests that the R-AGN incidence depends on either the properties (the mass) of the dark matter halo rather than the actual density of galaxies, or that BCG R-AGN have an enhanced availability of gas compared to R-AGN in the field due to the presence of a cooling flow \citep{2012ARA&A..50..455F}.
This flow may cause the AGN to always be in the ``on" state, in contrast to R-AGN in the field which are only intermittently triggered \citep{2014ARA&A..52..589H}. 
Despite the higher incidence rate of R-AGN in BCGs, the majority of R-AGN are found in the field.

We treat R-AGN with a low accretion rate, known as low-excitation radio galaxies (LERGs), separately from those with a high accretion rate, which are known as high-excitation radio galaxies (HERGs).
This is because they may not interact with the intergalactic medium in the same way. In addition to their radio jets it is possible that HERGs, which constitute only a few percent of the R-AGN population \citep{2012MNRAS.421.1569B}, also affect their host galaxies via AGN-driven winds. 
The LERGs and HERGs are distinguishable by their `excitation index,' which is derived from four optical emission lines \citep{2010A&A...509A...6B}.
Many previous studies have used the Fanaroff-Riley classification \citep{1974MNRAS.167P..31F} rather than the LERG/HERG separation. This classification uses radio morphology and radio luminosity to separate R-AGN into two categories. R-AGN that are core-brightened and of lower average luminosity are referred to as Fanaroff-Riley type 1 (FR1) sources, while R-AGN that are edge-brightened and of higher average higher radio luminosity are Fanaroff-Riley type 2 (FR2) sources. 
Although LERGs broadly overlap with FR1's and HERGs with FR2's, a portion of FR2 sources are LERGs \citep{1994ASPC...54..201L}. 
In our study we use the HERG/LERG classification of \citet{2012MNRAS.421.1569B} since it more closely matches the accretion rate and because the morphological information required to classify into FR types is generally not available for this sample.
As an alternative to excitation classification, we also split all R-AGN into groups according to what we call {\it specific radio luminosity} (SRL), the radio luminosity divided by the galaxy's stellar mass. We consider the normalized radio luminosity to be more relevant than just the radio luminosity in the context of investigating potential feedback because a jet of a given radio luminosity will probably not have the same effect on galaxies of different masses. We find that the SRL is better correlated with excitation classification than just the radio luminosity.

Radio jets are strongly anisotropic, and for feedback to be effective they must somehow deposit their energy to the surrounding halo gas. This is believed to occur as the expanding cavities and bubbles which the jets produce efficiently transfer the jet energy to the surrounding gas and disrupt the cooling of gas from the surrounding hot halo. 
This feedback mode could be responsible for maintaining the low star formation rates (SFRs) of early-type galaxies across cosmic time \citep{2003ARA&A..41..191M, 2006MNRAS.370..645B}, however there is a lack of evidence, not only for a causal connection between the presence of AGN and the suppressed SF, but even that the two phenomena are correlated. A given passive, or nearly passive galaxy can be passive for a number of different reasons, including e.g., the environment. The goal of this study is to search for such a correlation.
 
The strongest evidence for R-AGN heating of halo gas comes from X-ray observations of clusters. 
\citet{2003ApJ...590..207P} studied the X-ray spectra of 14 cooling flow clusters and found that the gas temperatures at the cluster cores are only $\sim$1/3 that at large radii, which is far below what is expected if a strong cooling flow were taking place. 
The Perseus cluster shows evidence of weak shocks and ripples in the X-ray, which could be evidence of the expanding radio bubbles heating the ambient gas \citep{2003MNRAS.344L..43F}.
X-ray bubbles may also be seen in individual elliptical galaxies. 
\citet{2007hvcg.conf..210N} studied a sample of 104 ellipticals and found cavities in 24 of them, which could indicate that the R-AGN duty cycle is lower in individual galaxies ($\sim$25\%) than it is in cluster environments, where it may approach 100\% \citep{2008MNRAS.385..757D}. 

A few studies have directly examined the effects of feedback on the host galaxy's SFRs, with conflicting results. 
\citet{2013ApJ...774...66Z} measured the far infrared (FIR) SFRs of a sample of $\sim$3,000 X-ray and radio-selected AGNs in the Chandra Deep Field South and found that R-AGN have a much higher SFR than purely X-ray selected AGNs, which appears to be at odds with the scenario of R-AGN quenching star formation in their hosts. 
The difference may arise from the \citet{2013ApJ...774...66Z} sample being from an earlier cosmic epoch ($z\sim2.0$).
The simple picture of R-AGN negatively affecting star formation is further complicated by the results of \citet{2014ApJ...784..137K}, who examined the infrared star formation rates of a sample of $\sim$300 radio sources in the North Ecliptic Pole (NEP) field and found a positive correlation between the luminosity of the AGN component and star formation, which suggests that greater AGN activity leads to increased SF. 
However when the sample is binned by redshift and AGN luminosity, the specific star formation rate (SSFR) is found to decrease with increasing radio luminosity. 

Interactions with radio jets may affect star formation in adjacent satellite galaxies. 
Powerful R-AGN may induce star formation in their satellies, as seen in the FR2 sources PKS 2250-41 \citep{2008MNRAS.386.1797I} and 3C 285 \citep{1993ApJ...414..563V, 2007ApJ...662..166H}.
Interactions with weak R-AGN may also induce star formation, as seen in the FR1 sources associated with Centaurus A \citep{1998ApJ...502..245G, 2012MNRAS.421.1603C} and Minkowski's Object \citep{2004IAUS..222..485V, 2006ApJ...647.1040C}.
On the other hand, radio jets may quench star formation in satellites.
\citet{2011MNRAS.413.2815S} found that the $u-r$ colors of satellite galaxies in the projected  path of FR2 jets are redder than those outside of the path, but no such trend was found for FR1 sources.
This may indicate that interactions with powerful radio jets quench star formation, while weaker jets are not capable of doing so. 
This is in contrast to the results of \citet{2014ApJ...785...66P}, who found no clear difference in the satellite colors of R-AGN and normal galaxies, regardless of the AGN accretion mechanism, which suggests that on average R-AGN have little influence on star formation in their hosts.

In this work, we will look for evidence of R-AGN feedback on their hosts by statistically exploring the relationship between R-AGN and recent star formation.  This will be done by comparing the composite spectral energy distributions (SEDs) of a sample of R-AGN to that of a carefully selected control sample. 
The composite SEDs are generated from UV, optical, and mid-IR photometry from the Galaxy Evolution Explorer (GALEX), the Sloan Digital Sky Survey (SDSS), and the Wide-field Infrared Survey Explorer (WISE), respectively.
Since many of the R-AGN and control galaxies have weak detections in the UV and mid-IR, we adopt a Bayesian approach to produce the composite SEDs, which removes the biases of only considering the detections above some S/N ratio threshold. 

This paper is organized as follows. Section 2 describes the data sources used as well as how this information was combined. Section 3 details the selection of the R-AGN and control samples while Section 4 describes our method of constructing the composite SEDs. Section 5 presents the results of our study, and in Sections 6 and 7 we discuss our findings. The cosmological parameters we have adopted in this work are $\Omega_m=0.3$, $\Omega_\Lambda=0.7$, and $H_0=70$ km s$^{-1}$ Mpc$^{-1}$.

\section{Construction of Samples}

Our analysis is based on a sample of R-AGN, drawn from \citet{2012MNRAS.421.1569B}, that have GALEX coverage to a uniform depth. 
Like the R-AGN sample, the control sample is drawn from areas of SDSS that are covered by the radio surveys as well by GALEX to the appropriate depth.
Because WISE is an all-sky survey, both the R-AGN and control galaxies are by definition covered by WISE.
The selection and properties of these samples is described below.

\subsection{R-AGN sample}
\label{data}

A sample of 18,286 radio-emitting galaxies  was created by Best \& Heckman 2012 (hereafter BH2012) by combining galaxies with spectra from the 7th SDSS data release (DR7; Abazajian et al. 2009) with two radio surveys: the National Radio Astronomy Observatory (NRAO) Very Large Array (VLA) Sky Survey (NVSS; Condon et al. 1998) and the Faint Images of the Radio Sky at Twenty centimeters (FIRST; Becker, White \& Helfand 1995) survey. 
Since our study focuses in the UV and mid-IR emission of the R-AGN, we consider the portion of the BH2012 sample that is covered by the Galaxy Evolution Explorer (GALEX; Martin et al. 2005) to a uniform depth.
We therefore selected BH2012 galaxies with GALEX exposure times in the range of $500<exptime<4,500$ seconds corresponding to what is known as the Medium-depth Imaging Survey (MIS). Although the range of exposure times is somewhat large, the majority of observations have exposure times of $\sim$1,500 s. This resulted in an initial pool of 5,400 galaxies.
 
Our study of these R-AGN used data from three large-area, multiwavelength surveys. Ultraviolet photometry were drawn from the DR6/DR7 GALEX data release, optical photometry were drawn from the tenth data release (DR10; Ahn et al. 2014) of the Sloan Digital Sky Survey (SDSS), and infrared photometry were drawn from the AllWise catalog (Wright et al. 2010, Mainzer et al. 2011). 
The GALEX survey is an all-sky survey in two UV bands: the far-ultraviolet band (FUV) is centered at a wavelength of 1528 \AA$\,$ while the near-ultraviolet (NUV) band is centered at a wavelength of 2271 \AA.
The WISE mission surveyed the entire sky in four bands, designated W1 through W4. The W1 band is centered at 3.3 $\mu$m, the W2 band at 4.6 $\mu$m, the W3 band at 12$\mu$m, and the W4 band at 22 $\mu$m. 
We required the control galaxies to have the same depth in GALEX as the R-AGN, and  found that there are 361,469 SDSS galaxies that are covered in GALEX with the required depth. Fluxes, magnitudes, and the DR10 reported redshifts were extracted from the three surveys for these galaxies. 
A nearest-neighbor approach was used to combine information from the surveys. We used a 5" radius to match the GALEX and WISE sources to the SDSS sources. Approximately 0.02\% of the R-AGN had multiple SDSS matches and 9\% had multiple WISE matches in the 5'' radius, and in these situations we kept the source with the smallest separation. Choosing the closest match in WISE limits the fraction of potential spurious matches to $<0
.1\%$.
 
The SDSS and GALEX photometry was used to derive the physical properties (stellar mass, mass-weighted ages, SFRs) of the galaxies via UV-optical SED fitting to stellar population synthesis models, which were then used in the selection of the control sample as well as the analysis. 
The fitting was performed following the methodologies of \citet{2007ApJS..173..267S} and \citet{2014ApJ...797..126S}. We perform the fitting by comparing the observed fluxes to a library of model fluxes, which are derived from stellar population synthesis models \citep{2003MNRAS.344.1000B}. Star formation histories are combinations of exponentially declining trends (with inverse decline rates sampled from 0 to 1 Gyr$^{-1}$), with uniformly distributed formation times, and superimposed stochastic bursts (with burst probability of 50\% over 2 Gyr). Metallicities take values uniformly from 0.02 to 2 Z$_{\odot}$. Details are given in \citet{2008MNRAS.388.1595D}. Stellar masses and SFRs were derived using Chabrier IMF. The \citet{2000ApJ...539..718C} dust extinction model is applied to model SEDs. Libraries of model fluxes are produced in redshift steps of
z = 0.01.

The initial pool of BH2012 galaxies contains R-AGN as well as galaxies in which the radio emission is not from a nuclear source, but is associated with SF (i.e., SN remnants, Condon 1992).
A combination of three methods were used by \cite{2012MNRAS.421.1569B} to separate the R-AGN from the normal galaxies in their sample. Two of these methods (4000 \AA~ break strength vs. ratio of radio luminosity per stellar mass; H$\alpha$ line luminosity vs. radio luminosity) are based on the assumption that R-AGN have enhanced radio emission compared to normal galaxies with the same SF, where $D_{4000}$ and H$\alpha$ serve as proxies for the latter. 
The third method, which is the emission-line diagnostic or `BPT' diagram (Baldwin et al. 1981, Kauffmann et al. 2003),  relies on differences in excitation due to the relative strengths of the ionizing radiation of AGN and O stars. While these separation methods usually agree, they may occasionally give contradictory classifications. For a full description of how a final classification was determined in such cases, see Appendix A of BH2012.

\begin{figure}[t!]
\includegraphics[width=3.51in]{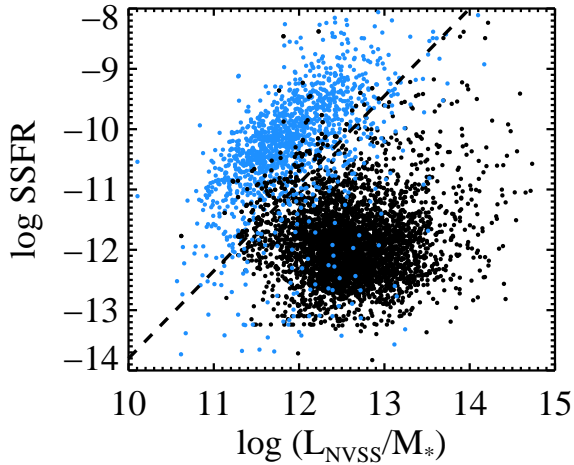}
\caption{ Plot of $log(SSFR)$ vs. $log(L_{NVSS}/M_\star)$ for the 5,400 galaxies \citet{2012MNRAS.421.1569B} galaxies that are covered in the GALEX MIS. The dashed line shows our division between star-forming galaxies and R-AGN, while the galaxies classified by \citet{2012MNRAS.421.1569B}  as star-forming and R-AGN are shown as blue and black points, respectively. Those R-AGN (black points) that lie above the line are plotted on top of the star-forming galaxies (blue points) while below the line the star-forming galaxies are plotted on top of the R-AGN. In most cases our classification agrees with that of \citet{2012MNRAS.421.1569B}.
}
\label{ssfr}
\end{figure}

The H$\alpha$ line luminosity is an indicator of star formation, but is affected by dust attenuation and AGN emission. Furthermore the observed H$\alpha$ luminosities in the SDSS database are not corrected for the fixed fiber sizes of the SDSS spectrograph. No dust or aperture correction was applied by BH2012, so it is possible that the H$\alpha$ emission of some of the R-AGN may not reflect their true star formation rate.
SED-based SFRs similar to those used here were already subjected to extensive comparison with H$\alpha$ based SFRs in Salim et al. (2007). It was found that for galaxies with low levels of SF the SED based SFRs are superior to those based on H$\alpha$.
The specific star formation rates (SSFRs), which are the SFRs divided by the stellar mass, from our SED fitting provide another measure of star formation activity, and can be used to filter out any galaxies for whom the H$\alpha$ luminosity is a poor indicator of the true star formation rate.
Figure~\ref{ssfr} shows the distribution of BH2012 sources in the log (SSFR) vs. log($L_{radio}/M_\star$) plane. The sample is clearly composed of two types of galaxies: those with high SSFRs but low radio luminosity per stellar mass (upper left) and those with a low SSFR but a high radio luminosity per stellar mass (bottom). The dashed line at 
\begin{equation}
\log (SSFR)=1.45\times\log(L_{radio}/M_\star)-28.3,
\end{equation}
shows our division between star-forming galaxies and R-AGN. Galaxies classified as star-forming and R-AGN by BH2012 are shown in blue and black, respectively. 
Most of the BH2012 star forming galaxies lie above the line while most of the galaxies that BH2012 classify as R-AGN lie below the line. 

\begin{deluxetable}{rcccc}[h!]
\tabletypesize{\scriptsize}
 \tablecolumns{9}

 \tablecaption{Number of R-AGN per accretion mode and redshift bin for each environment type. \label{tbl1}}
 \tablehead{
    \colhead{Environment}  & \colhead{HERG} & \colhead{LERG} & \colhead{Unclassified}  & \colhead{Total}  }

 \startdata
  Field&108&2,536&446&3,090 \\

Member&4&201&33&238 \\  

BCG &3&421&83&507\\
\tableline
\rule{0pt}{3ex}
Total&115&3,158&562&3,835\\

\enddata
\tablecomments{ Columns 2, 3, and 4 are the accretion modes as defined by \citet{2012MNRAS.421.1569B} and column 5 is the number of R-AGN per environment type. }
\label{mytable}
\end{deluxetable}

For our sample we retain those BH2012 R-AGN (black points) that lie below the line. 
The SDSS data pipeline assigns objects a category of either {\tt GALAXY, STAR,} or {\tt QSO} based on a comparison of the object's spectrum to a range of template spectra. None of the R-AGN below our dividing line were classified by the SDSS pipeline as {\tt STAR}. The UV emission of quasars will include a contribution from the AGN, which would affect our ability to use UV as a indicator of SFR. LERGs lack an accretion disk and should therefore have weak or no UV emission while the UV from the accretion disks of HERGs is blocked by the dusty obscuring structure, so any UV observed in R-AGN should trace star formation.
We therefore excluded 65 R-AGN that are classified by the SDSS DR10 pipeline as quasars, leaving a final sample of 3,835 R-AGN.
The magnitude range of this sample is 11.72 $< r <$ 18.62, with a redshift range of 0.01 $< z <$ 0.3. The lower redshift limit is that of the original B\&H sample, and the upper redshift limit that we have adopted matches that of the SDSS main galaxy sample \citep{2002AJ....124.1810S}, thus providing a large pool of candidate control galaxies.

We assign an environment classification for each galaxy in this sample, following the procedure of \citet{2014ApJ...785...66P}. Three environment classifications are used: field galaxy, cluster member, and BCG.
The DR10 coordinates and redshifts of the R-AGN sample were used to calculate the projected distance to the nearest BCG in the \citet{2010ApJS..191..254H} catalog of BCGs. Those R-AGN with a projected distance to the nearest BCG of less than 1.5 kpc were classified as BCGs, while those with a projected distance between 1.5 kpc and 1.5 Mpc and a difference in spectroscopic redshift less than $z=0.01$ were classified as cluster members. The remaining galaxies were classified as field galaxies. The number of galaxies in each environment are shown in Table~\ref{tbl1}.

Because we want to address how the accretion mechanism relates to feedback effects, we have used the LERG and HERG classifications provided by BH2012. These classifications are based on the ``excitation index" defined by \citet{2010A&A...509A...6B} and in most cases (84\% of the sample) can be used to classify the galaxy as either a HERG or LERG. 
Many of the remainder could not be classified because their emission lines are undetected, which could because they are LERGs, but could also be due to their faintness.
Table~\ref{tbl1} shows that 115 or 3\% of the sample are HERGs, while 3,158 or 81\% of the sample are LERGs.  Table 1 shows that HERGs are almost never found in high-density environments. Most LERGs are also found in the field, although they are more likely than HERGs to be found in cluster members or BCGs.

\subsection{Control sample (non R-AGN)}

We draw the control sample from a pool of 331,145 candidate galaxies, which, like our R-AGN sample, are covered by GALEX observations of intermediate depth, and fall within the FIRST footprint. The latter is essential to be able to establish that the galaxy is not an R-AGN. For each R-AGN we identify one control galaxy, by matching by environment (i.e., being in the field, in a cluster, or a BCG) and by redshift, stellar mass, population age, and the axis ratio. The purpose of these matching criteria is to ensure that the control sample has the same quality of photometry as the R-AGN sample (ensured by redshift matching) and that the galaxies have similar evolutionary histories (matching by stellar mass and mass-weighted population age). Matching by population age ensures that the galaxies have roughly similar SF histories. The population age, being mass-weighted and therefore dominated by past SF, is not sensitive to the current SFR, which is what we wish to compare between R-AGN and the control sample. 
Because most radio galaxies are ellipticals (which tend to be rounder), we also apply matching by axis ratio in order to reduce the occurrence of lenticulars (S0s), which tend to be flatter, among the control galaxies.  
Stellar masses and population ages of S0s may overlap with those of the ellipticals, but S0s can nevertheless have higher current levels of SF \citep{2012ApJ...755..105S}. 

  \begin{figure}[t!]
\includegraphics[width=3in]{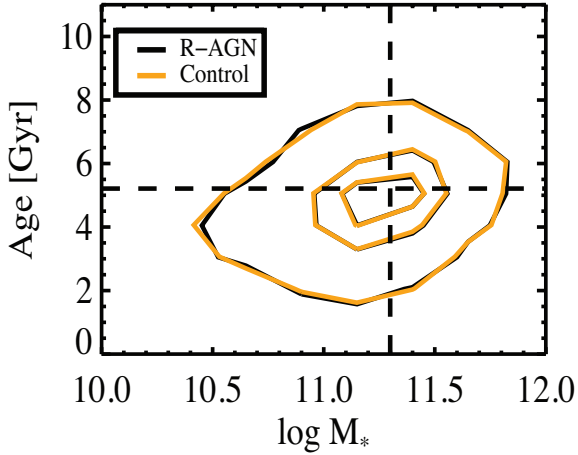}
\caption{Comparison of the R-AGN and control samples in the stellar mass versus stellar population age plane. The contours enclose 90\%, 50\%, and 25\% of the galaxies in the samples. The dashed lines denote the division between old and young and high and low mass populations, discussed in Section 2. This figure shows that the control galaxies have been successfully matched to the R-AGN.}
\label{control}
\end{figure}

For each R-AGN, candidate control galaxies were first selected to have the same environment type (field candidate for field R-AGN, etc.), then the best match from among these candidates was chosen by minimizing the metric 
\begin{equation}
 \small
 R \,= \,\sqrt{\left(\frac{\Delta z}{0.24}\right)^2+\left(\frac{\Delta\log M_\star}{1.23}\right)^2+\left(\frac{\Delta age}{6.00}\right)^2+\left(\frac{\Delta( b/a)}{0.55}\right)^2}.  
 \end{equation} 
Here the quantities $\Delta z$, $\Delta\log M_\star$ [$M_\sun$], $\Delta age$ [Gyr], and $\Delta(b/a)$ represent the difference between the R-AGN and control candidates in redshift, stellar mass,  population age, and axis ratio, respectively. The denominators represent the 95 percentile range of these properties for the R-AGN sample and are used so that all criteria have the same relative weight.
This selection process sometimes results in one control galaxy being the best match to more than one R-AGN. This occurs for 9\% of the controls and has no effect on our analysis.

Figure~\ref{control} shows the age and stellar masses of the control sample, plotted together with those of the R-AGN. The contours enclose 90\%, 50\%, and 25\% of the samples. Divisions in age and mass indicated by the dashed lines are used in the analysis in Section 4. This plot illustrates the success of our matching process, as galaxies in the control sample have the same distribution as the R-AGN in age as well as stellar mass.

\section{Methods}
\label{methods} 

The analysis is based on the comparison of aggregate SEDs of R-AGN vs. the SEDs of control galaxies. Each comparison set is a part of the full sample binned or classified by some property. For example, we may wish to compare the SEDs of BCG R-AGN to BCG control galaxies. An SED consists of luminosities, or absolute magnitudes, at 11 wavelengths (2 UV, 5 optical and 4 mid-IR). Galaxies in our sample lie at a range of redshifts ($0.01<z<0.3$) so the observed fluxes sample SEDs at a range of rest-frame wavelengths. To obtain comparable SEDs we construct  them as if all galaxies were at $z=0.1$. To achieve this we assume that small sections of SEDs can be described with a slope $\alpha_\lambda = dM_\lambda/d (\log \lambda)$. Therefore, to construct the SED we need to obtain 11 M$_\lambda^{z=0.1}$ and $\alpha_\lambda$.

The standard approach for obtaining each of the 11 SED points would be to correct the observed fluxes of galaxies in the comparison set for Galactic reddening and then convert the fluxes into absolute magnitudes, to which M$_\lambda^{0.1}$ and $\alpha_\lambda$ are fitted. However, this procedure would be problematic for galaxies with low S/N measurements, which may sometimes even have negative measured fluxes. This is the case for many of the galaxies in our sample, usually more distant galaxies in the far UV and at 12 and 22 $\mu$m. Rather than introducing some S/N threshold which will bias M$_\lambda^{0.1}$ towards stronger detections, we take a Bayesian approach to determine the SED, i.e., we ask what M$_\lambda^{0.1}$ and $\alpha_\lambda$ can best explain the observed fluxes of galaxies in the comparison set. Therefore, we assume some M$_\lambda^{0.1}$ and $\alpha_\lambda$ and predict the observed fluxes based on the distance to the galaxy and the Galalactic reddening for that galaxy. 
We try different values of  M$_\lambda^{0.1}$ and $\alpha_\lambda$ until the $\chi^2$ 
\begin{equation}
 \chi^2 = \sum \frac{(flux_{model}-flux_{obs})^2}{\sigma_{flux, \,obs}^2}, 
\end{equation}
is minimized.
Predicting the observed flux from model absolute magnitude requires an additional correction. Namely, the galaxies in a comparison set have a range of stellar masses and the SEDs will to first order differ simply because of the scaling with the mass. To remove this overall variation we construct the SED assuming all galaxies had $\log M^*=11.3$, the median mass of the sample. 
Therefore, in order to produce a predicted flux, the model absolute magnitude is adjusted by 
\begin{equation}
\Delta M_\lambda = -2.5\times  (\log M^*-11.3).
\end{equation}

To test the robustness of our technique, we perform the simulation of the Bayesian estimation of the absolute magnitude in W4 (22 $\mu$m), the band where the detections are the weakest. We assume that the true absolute magnitude in W4 at $z=0.1$ is -22.0 (for runs \#1 and \#2) and -21.5 (for run \#3). These assumed values bracket the actual absolute magnitudes at 22 $\mu$m. We produce 1000 (runs \#1 and \#3) and 100 (run \#2) such objects and distribute them over the redshift range $0.04<z<0.3$, peaking at $z=0.15$, similar to the actual redshift distribution. The simulated sample size (100--1000) brackets the numbers in the actual analysis. We also assume the slope of the SED to be $-20$, a typical true value. Based on the assumed absolute magnitude at $z=0.1$, the assumed SED slope and the redshift, we calculate for each object its true flux and, based on it and the sensitivity of WISE, the expected flux error. We add an additional 2\% error to mimic calibration uncertainties. Finally, we perturb true fluxes with the value drawn from a Gaussian with the standard deviation equal to the flux error. We then use these simulated fluxes to reconstruct $M_{W4}^{0.1}$ and $\alpha_{W4}$ using the Bayesian approach described above.

  \begin{figure}[t!]
\includegraphics[width=3.5in]{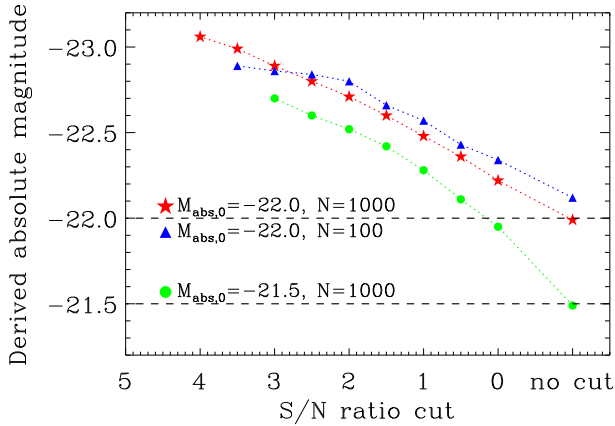}
\caption{The effects of S/N cuts on Bayesian estimates of absolute magnitudes. Three simulations of differing sample sizes and initial absolute magnitudes (horizontal dashed lines) are shown. Each simulation was repeated multiple times, first with no S/N cut which produced a derived absolute magnitude that closely matched the input magnitude (rightmost points). Progressively more stringent S/N cuts resulted in the derived absolute magnitude being increasingly brighter than the true absolute magnitude.}
\label{simres}
\end{figure}

For runs with 1000 objects we retrieve the absolute magnitude within 0.01 mag and slope within 2 (10\%). For the run with 100 objects the estimated magnitude is 0.1 mag off from the actual value. These results are shown as the rightmost points in Figure \ref{simres}. This figure also shows what results we would have obtained if we used a standard approach in which only the detections above some signal-to-noise (S/N) ratio are used in the calculation. Because W4 detections are weak, only a small percent of the objects reach S/N$>3$. Absolute magnitudes that would be obtained only using these measurements are 1 to 1.5 magnitudes brighter than the actual value, and the slopes are much steeper than what they ought to be. Lowering the S/N ratio cut reduces the discrepancy, but the derived absolute magnitudes are still too high (0.3 - 0.5 mag) even when all objects with positive fluxes are used (S/N$>$0). The least biased, and, as we show, rather accurate, answer is obtained only when all of the data are used, including those having ``unphysical'' negative fluxes, which comprise 20--30\% of the simulated samples.

If the galaxies in the comparison set were truly homogeneous or drawn from a normal distribution the above procedure would be sufficient to obtain a typical SED: 11 M$_\lambda^{0.1}$ and their errors. The knowledge of robust errors is critical to be able to establish if the SEDs of R-AGN and the control sample differ significantly or not. However, we have noticed that in the UV and 12 and 22 $\mu$m there is a tail of galaxies with higher fluxes than that of the bulk of the galaxies in a given comparison set. 
They drive the minimum $\chi^2$ to high values and render the absolute magnitude error from the $\chi^2$ analysis useless (because it assumes the underlying distribution of fluxes to be normal). Furthermore, the best-fit (minimum $\chi^2$) absolute magnitude will implicitly ignore these outliers, whose presence is important for understanding the properties of the comparison set. We address these issues by determining M$_\lambda^{0.1}$ and its error in a way that is more appropriate given the circumstances. 
Namely, we take the SED slope derived from the $\chi^2$ minimization, but now look for M$_\lambda^{0.1}$ such that 50\% of the observed fluxes will be higher than the predicted value and 50\% lower, i.e., we obtain M$_\lambda^{0.1}$ that is the equivalent of a median, rather than the weighted average as in the case of $\chi^2$ minimization. 
Note that we used the information on the SED slope from the $\chi^2$ analysis, as it can not be determined by this ``median" estimate. To determine the error of  M$_\lambda^{0.1}$ we perform 100 determinations of the median from bootstrap resampled data points and determine the 84th and 16th percentile ranges of such medians. Thus we obtain separate upper and lower error bars. This error takes into account not only the flux errors but also the range due to the intrinsic distribution of absolute magnitudes in the comparison set. For all flux points the errors of magnitudes from the aggregate SED is dominated by the range of values of individual galaxies in the sample, not the photometric error.

\begin{figure*}[t!]
  \centering
\includegraphics[width=5.8in]{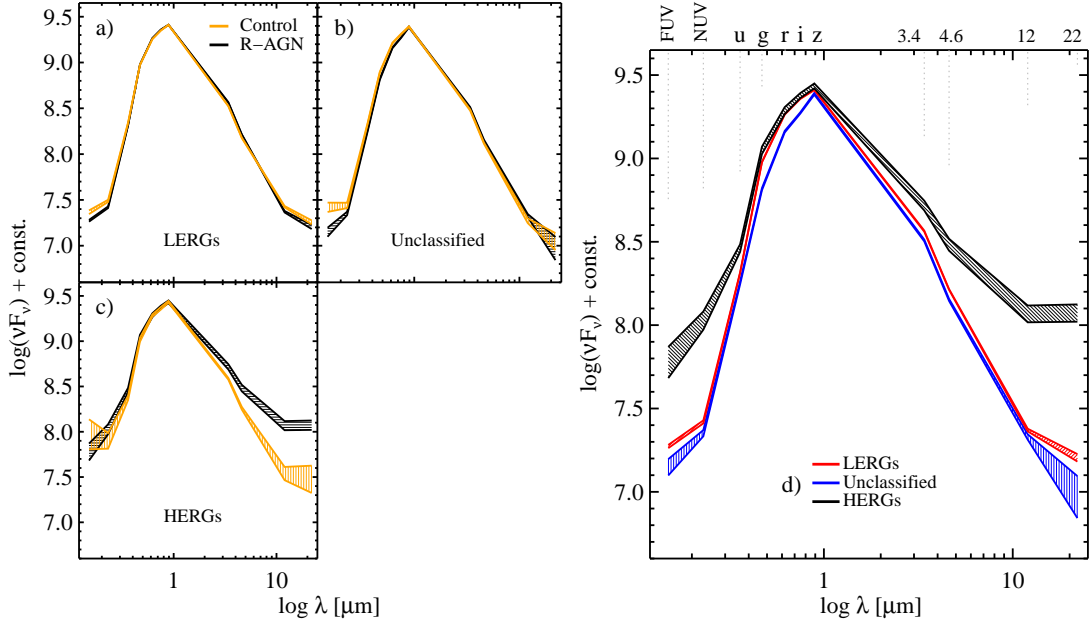}
\caption{The SEDs of the R-AGN and their controls, separated by accretion mechanism. The LERGs, in panel (a), have suppressed UV and mid-IR relative to their controls, while the unclassified R-AGN in panel (b) show suppressed UV while the mid-IR emission is similar to that of the controls. The UV of the HERGs in panel (c) is similar to that of their controls, while their mid-IR is much stronger. The right-hand panel compares the different accretion modes and shows that those R-AGN that are unclassified have UV and mid-IR emission similar to though at lower levels than that of the LERGs, while that of the HERGs is strongly enhanced. }
\label{accretion}
\end{figure*}

\section{Results}
In this paper, we study the possible effects of R-AGN on their host galaxies by comparing the SEDs (especially the UV and mid-IR portions) of the R-AGN and control samples. We split the sample into comparison sets by accretion mode, specific radio luminosity, environment, and mass and age.
In addition to comparing the R-AGN and control galaxies, we also compare the SEDs of the R-AGN, binned according to these properties. 
We also evaluate the effectiveness of mid-IR colors at selecting R-AGN of either accretion type as well as how mid-IR colors are related to the star formation rates and radio luminosity of the R-AGN and the host galaxy.

\subsection{Accretion mechanism}
We first consider how differences in the accretion mechanism are reflected in the composite SEDs. 
For this portion of the analysis we restrict ourselves to those galaxies in the field, which make up the majority of the sample (Table~\ref{mytable}). 
This is done because the large-scale environment may have an effect on the SEDs, and we explore these effects separately. 
Panels (a)-(c) of Figure~\ref{accretion} compare the SEDs of the R-AGN to those of their controls, while panel (d) shows only the R-AGN. Panel (a) shows that LERGs have less UV and mid-IR emission  than their controls, while only the UV emission of the unclassified R-AGN in panel (b) is less than that of their controls. In contrast, the UV emission of the HERGs in panel (c) is similar to that of their controls, while the mid-IR of the HERGs is greatly elevated.
The comparison of the R-AGN with different accretion modes, in panel (d) of Figure~\ref{accretion}, shows that the overall shape of the SEDs for the LERG and unclassified R-AGN are more similar to each other than compared to the HERGs. 
Both have UV luminosities that correspond to relatively passive galaxies ($\log SSFR\sim12$), 1.5 dex lower than the SFR of galaxies on the main sequence having the same mass.
However, the SEDs of LERGs and unclassified R-AGN exhibit some differences. Most notably, unclassified R-AGN  are fainter in $g$, $r$ and $i$. This is consistent with them being on average older (0.2 Gyr) and more massive (0.11 dex) than the LERGs. 
This likely a result of the ratio of unclassified R-AGN to LERGs increasing from $<1\%$ at $z=0.1$ to $>50\%$ at $0.28<z<0.3$, such that in the higher redshift bins the mean luminosity of the unclassified R-AGN is greater than that of the LERGs (Malmquist bias).

Moving on to the HERGs we see that their optical emission is quite similar to that of the LERGs, but their UV and mid-IR emission is strongly elevated (Figure~\ref{accretion} d). 
This difference is greatest in the UV and longest two WISE bands, although it is seen in the other two WISE bands as well.
HERGs are therefore consistent with having higher levels of SF than LERGs and/or higher levels of dust heating from an AGN.
The specific SFR of HERGs is $\log SSFR\sim11$, which is 0.6 dex higher than that of LERGs.

Panels (a)-(c)  show the SEDs of the R-AGN and their associated controls in the three accretion mechanism categories. 
Both the UV and mid-IR emission of the LERGs in panel (a) is lower (FUV: 0.09 dex; NUV: 0.07 dex; 12 $\mu$m: 0.06 dex; 22 $\mu$m: 0.05 dex) than that of the control galaxies, although the difference at 22 $\mu$m is not statistically significant. The R-AGN for which BH2012 provide no classification are shown in panel (b), and here the UV is lower for the R-AGN, while emission in the WISE bands is broadly similar to that of the controls.
The HERG sample, shown in panel (c), only has $\sim$100 members which results in greater uncertainty in the composite SEDs for both the R-AGN and controls. Even so, there are clear differences between the two. The HERGs have a very strong excess across all four mid-IR bands. Because the UV emission of HERGs and their controls is similar, all of the mid-IR excess can be attributed to dust heating from the AGN. 

These results support the picture of HERGs being found in galaxies with sufficient cold gas to fuel both radiatively-efficient accretion, which is manifest as mid-IR emission from heating of the obscuring torus, and star formation, which is manifest as elevated UV emission.
The fact that we see similar levels of UV emission in both HERGs and their controls is still consistent with this idea because not every galaxy with a large quantity of cold gas will be an R-AGN, but when it is, it will be a HERG. 
The AGN and star formation activity appear to have little effect on one another, which is evident in the similarity of the UV portions of the composite SEDs in panel (c). The weak mid-IR signature of the LERGs probably results from their lack of an accretion disk and associated obscuring torus, while their weak UV emission reflects the low star formation rates of their hosts, and is consistent with the idea that their hosts have little available cold gas. However, unlike the HERGs the fact that the LERGs have both suppressed UV and IR emission suggests that there is a link between star formation LERG activity.

  \begin{figure*}[t!]
  \centering
\includegraphics[width=5.8in]{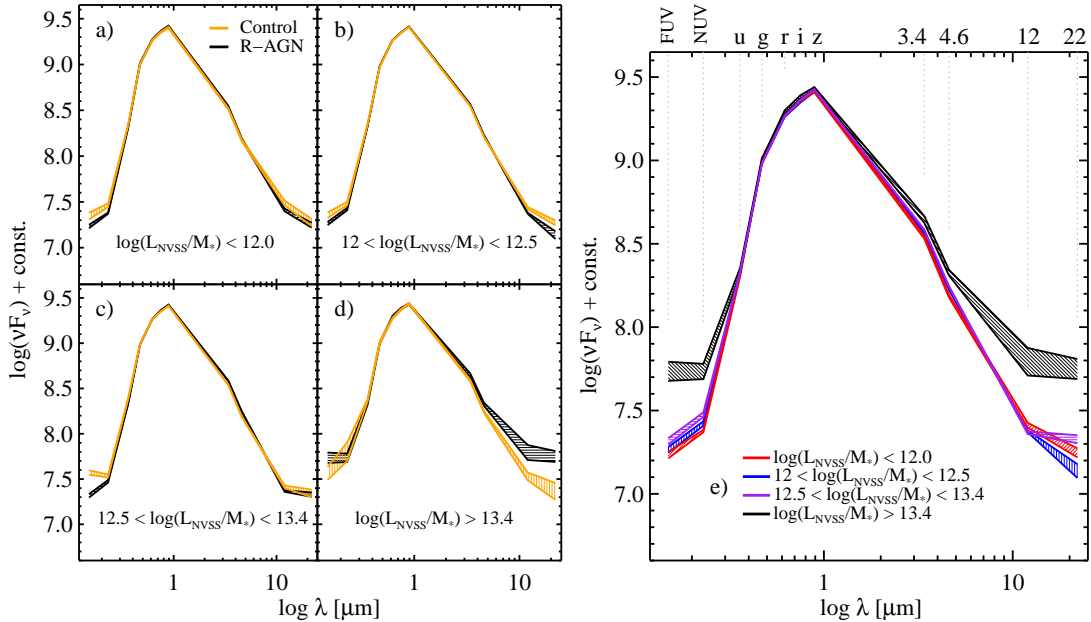}
\caption{The specific radio luminosity of field R-AGN, divided into four bins. The R-AGN in the first three bins, shown in panels (a)-(c), Have a deficit of UV eimission, while their mid-IR emission is similar to that of their controls. This suggests that the R-AGN in these bins are mostly LERGs, and they may suppress star formation in their hosts. The R-AGN in panel (d) have an excess of mid-IR emission, indicating that many of them are HERGs. Panel (e) compares the R-AGN in the four SRL bins and shows that those in the bin with the highest SRL have enhanced UV and mid-IR emission compared to the R-AGN in the other three bins.}
\label{radio}
\end{figure*}

\subsection{Specific radio luminosity}

Next we consider an alternative way to classify R-AGN. Not by excitation classification, but by the actual radio power, i.e., the specific (mass-normalized) radio luminosity. This was used together with the D$_{4000}$ break strength by \citet{2005MNRAS.362....9B} and \citet{2012MNRAS.421.1569B} to separate star forming galaxies from R-AGN, but here we consider its utility in separating high and low-accretion rate R-AGN. The radio luminosities are expressed in W Hz$^{-1}$  and are calculated from the radio fluxes in BH2012. We divide all field R-AGN into four categories by SRL.

Panels (a)-(d) of Figure~\ref{radio} show the SEDs of the field R-AGN and their controls in four SRL bins, while panel (e) shows the comparison among the R-AGN in the different SRL bins. 
There is a trend of increasing UV emission with increasing SRL in the SEDs of the three bins with the lower SRL in panel (e). Although the trend is weak, as all three bins are within $\sim$0.1 dex, it is significant, as the derived error bands do not overlap. This trend may suggest that there could be a link between the fueling of low levels of SF and fueling weak R-AGN. Note that having a link, or a common cause of these two processes does not eliminate the possibility that the SF in R-AGN is nevertheless suppressed compared to galaxies without an AGN.
The bin with the highest SRL is very different, with a far greater level of both UV and mid-IR emission. The elevated emission in this bin is observed in all WISE bands, with the strongest difference being found in the longer two wavelength bands. 
The SED of the R-AGN with the highest SRL is similar to that of the HERGs, even though not all HERGs belong to this SRL bin ($\sim$1/3 do) nor are all R-AGN in this SRL bin HERGs ($\sim$1/3 are).

Panels (a) and (b) show the composite SEDs of the R-AGN and control galaxies in the two bins with the lower SRL.
In the log(L$_{NVSS}$/M$_*$)$<12.0$ bin the R-AGN have less UV and mid-IR emission (FUV: 0.12 dex, NUV: 0.08 dex; 12 $\mu$m: 0.07 dex), as do the R-AGN in the $12<$ log(L$_{NVSS}$/M$_*$)$<12.5$ bin (FUV: 0.1 dex; NUV: 0.06 dex; 12 $\mu$m: 0.06 dex; 22 $\mu$m: 0.12 dex).
These deficits in UV and mid-IR suggest that the R-AGN in these bins have lower SFRs than their controls.
This begins to change for the R-AGN in the next SRL bin in panel (c), where the mid-IR emission is now very similar to that of their controls, while their UV emission remains lower than that of the control galaxies. 
A possible explanation may be that in the mid-IR we are beginning to see AGN emission, countering the 12 and 22 $\mu$m deficit seen in the lower two SRL bins. The control galaxies in this panel exhibit a statistically significant upturn in UV emission, but its origin is unknown, as it does not appear in the other panels.
The trend of increasing mid-IR emission with increasing SRL becomes evident in panel (d), with the emission of the R-AGN being strongly elevated relative to that of the controls. The larger error bands for this bin are due to its small size, as it only has $\sim$200 R-AGN.
 
The results from this section are qualitatively similar to those from the previous section where we split by excitation. We see that the UV and mid-IR deficits are characteristic of weaker R-AGN, but the mid-IR deficit turns into a strong excess as the radio power increases.

\begin{figure*}[t!]
  \centering
\includegraphics[width=5.8in]{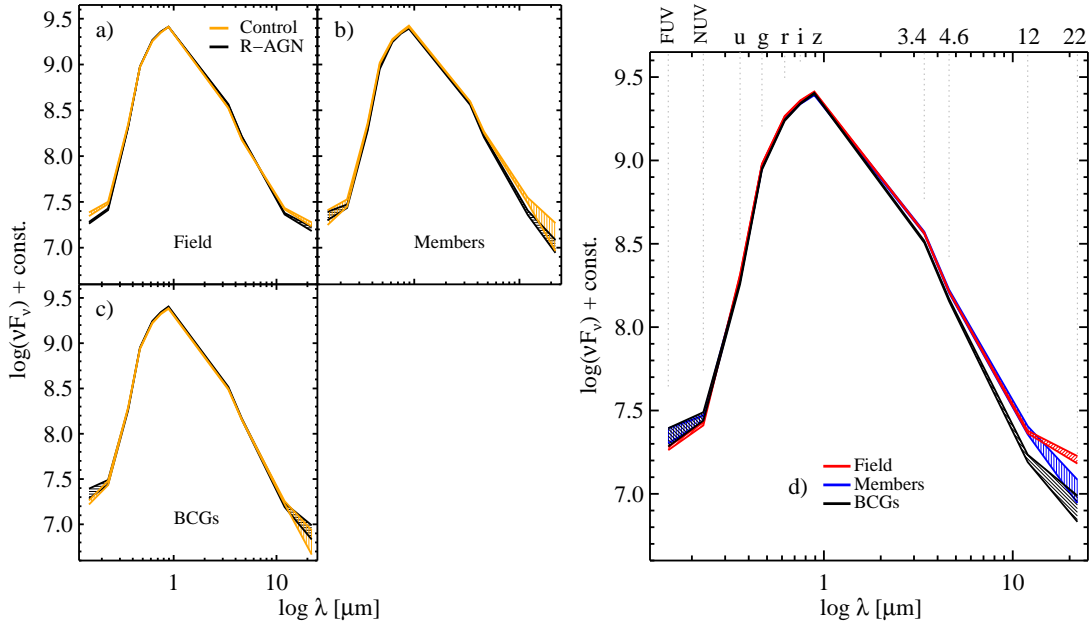}
\caption{ Environment effects on SEDs. The layout of this figure is as in Figure~\ref{agemass}, and is only for the LERGs, with those in the field in panel (a), cluster members in panel (b), and BCGs in panel (c). The feedback effects of LERGs do not appear to differ across environments. If we consider only R-AGN, the right-hand panel shows no difference in the UV portion of the SEDs, but there is a clear progression from lower to higher emission as the density increases. This may be a result of dusty hosts preferring sparser environments. }
\label{enviro}
\end{figure*}

\subsection{Environment}

We next explore differences in the SEDs for the R-AGN found in different environments: (1) BCGs, (2) cluster member other than BCG and (3) non-cluster galaxy (``field"). 
To isolate the effects of the environment, we analyze only LERGs, which make up the majority of the R-AGN (82\%) and are found in all three types of environment.
Panels (a)-(c) of Figure~\ref{enviro} show the SEDs of the LERGs and their controls in the three environments, while panel (d) shows the comparison of the SEDs for LERGs in the three environments. The optical and UV portions of the SEDs in panel (d) are very similar i.e., being in a cluster does not lead to a further decrease in the SFR. In the longest two WISE bands the emission is decreasing as the density increases. The separation here is much stronger, with the BCGs having the weakest emission in all four WISE bands. 
The slope of the SED at the longest two WISE bands may also change with density, largely due to weaker 12 $\mu$m emission in R-AGN BCGs. 
We also see that the BCG controls in panel (c) have weak 12 $\mu$m emission compared to the controls in the other environments.
These differences in the mid-IR are not due to differences in the SRL or population age, which are similar for all three categories. A possible explanation could be that BCG LERGs have much less dust than field LERGs. The W3 filter is also sensitive to PAH emission, so another explanation could be that PAHs are more easily destroyed in denser environments. 

Panel (a) shows the results for the field LERGs, which of course is the same as panel (a) of Fig.~\ref{accretion}.
As discussed above, the field LERGs show a slight deficit of mid-IR emission, while the UV is more strongly suppressed relative to that of the controls. Since the field LERGs is the largest sample their composite SEDs have the narrowest 1-sigma ranges so the differences between the LERGs and controls are the most robust of all the environments. 
There are only $\sim$200 cluster members, which explains the larger 1-sigma error bands in panel (b).The R-AGN and controls in this panel completely overlap in the UV, which suggests that being in a cluster can have a similar effect on SF as having an AGN.
However, other factors such as having a more massive dark matter halo or a smaller cold gas reservoir could also affect the SFR.
There is 0.06 dex deficit the 12 $\mu$m emission in the cluster member LERGs, while there are no significant differences in the SEDs of R-AGN BCGs and control BCGs, shown in panel (c). Like the R-AGN BCGs, the control BCGs have exceptionally low mid-IR emission, especially at 12 and 22 $\mu$m, consistent with the aforementioned possibility that BCGs are less dusty than the early type galaxies in other environments, which may be because the dust destruction timescales are shorter in the cores of galaxy clusters \citep{2010A&A...518L..47E}.

Overall, the broad similarity of the R-AGN SEDs in different environments suggests that the environment does not have an additional effect on SF.
 However, it is possible that the environment affects the amount of dust in galaxies, which could explain the trend for galaxies in denser environments having less mid-IR emission. In future studies we will investigate whether radio jets suppress star formation in LERGs in other environments, such as in groups and filaments.

  \begin{figure*}[t!]
  \centering
\includegraphics[width=5.8in]{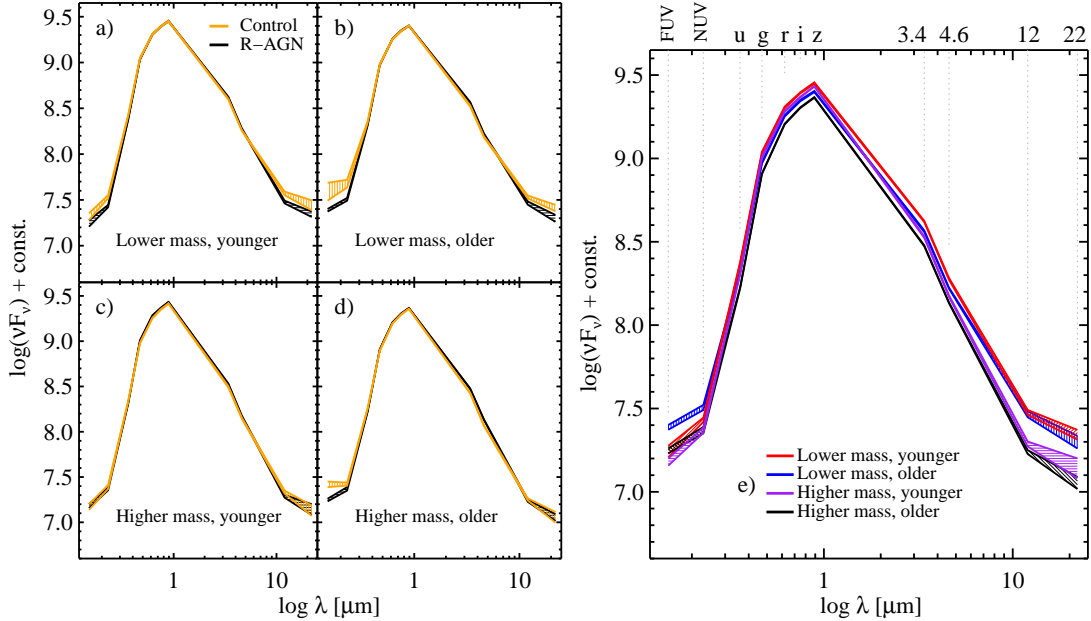}
\caption{The SEDs of field LERGs as a function of stellar mass and population age. The left panels show the results for R-AGN and control galaxies in four age and mass bins, split at $age = 5.2\,\,$ Gyr and $\log M^*= 11.3$,  while the right panel shows just the R-AGN in these bins. The lower-mass LERGs in panel (a) have less UV and MIR emission than their controls (FUV: 0.09 dex; NUV: 0.1 dex; 12 $\mu$m: 0.09 dex; 22 $\mu$m: 0.09 dex), as do the LERGs in panel (b) (FUV: 0.17 dex; NUV: 0.17 dex; 12 $\mu$m: 0.06 dex; 22 $\mu$m: 0.11 dex). Panel (e) shows that lower-mass LERGs have enhanced mid-IR emission with respect to higher-mass LERGs.}
\label{agemass}
\end{figure*}

\subsection{Mass and population age}
We also examined how the SEDs of R-AGN differ as a function of the mass and mass-weighted age of the host galaxy. For this portion of the analysis the effects of environment and accretion mode are controlled for by only considering LERGs in the field, which comprise 66\% of the sample. 
We have divided the field LERGs into two mass and two age bins, which allows us to compare the SEDs of galaxies with similar evolutionary histories such that the UV and mid-IR emission should be largely independent of these two properties.  The sample is split by the median mass and  age, at $age = 5.2\,\,$ Gyr and $\log M^*= 11.3$, and these divisions are illustrated by the dashed lines in Fig.~\ref{control}. 

Panels (a)-(d) of Figure~\ref{agemass} show the SEDs of the field LERGs and their controls in the four mass and age bins, and are a more detailed examination of the field sample shown in panel (a) of Fig.~\ref{enviro}.
Panel (e) shows the SEDs of the LERGs in the four bins. 
As described in Section~\ref{methods}, the SEDs are scaled by the median $\log M_\star$ of each bin, so the differences we see between galaxies in different mass bins are not due to the scaling of fluxes with mass, but rather reflect the differences in mass-to-light ratios.
The overall shapes of the SEDs in panel (e) are similar, although the galaxies with on average younger populations have stronger emission in the optical and 3.4 and 4.5 $\mu$m bands than the galaxies with populations that are on average older. This is because the mass to light ratio of the galaxies with a younger age is lower than that of the galaxies with ages that are older. We see a dichotomy in the IR, as the lower-mass galaxies have stronger emission in this portion of their SEDs while the difference due to age is less. The same dichotomy is seen for the control galaxies, which is consistent with more massive early type galaxies, whether AGN or not, being less dusty \citep{2012ApJ...748..123S}.
 The situation is less clear in the UV, as the effects of mass and age may be confounded.

Previously we have seen that field LERGs on average have lower UV and mid-IR emission at 12 and 22 $\mu$m than normal galaxies. We can now see which part of the sample contributes the most to this deficit. In panels (a) and (b) concurrent UV and mid-IR deficits are seen in lower mass LERGs, regardless of the population age. The differences for galaxies in panel (a) are FUV: 0.09 dex; NUV: 0.1 dex; 12 $\mu$m: 0.09 dex; 22 $\mu$m: 0.09 dex, and for panel (b) they are FUV: 0.17 dex; NUV: 0.17 dex; 12 $\mu$m: 0.06 dex; 22 $\mu$m: 0.11 dex.
While the SEDs of the R-AGN and controls in panel (c) are quite similar, a UV deficit may be present for the LERGs in panel (d) as well. The mid-IR looks the same for the R-AGN and controls, possibly because the emission is already at the minimum because of little dust and therefore the differences in SF are not reflected there.

In future studies we will investigate whether radio jets suppress star formation in LERGs in other environments, such as in groups and filaments.

  \begin{figure*}[t!]
  \centering
\includegraphics[width=5.8in]{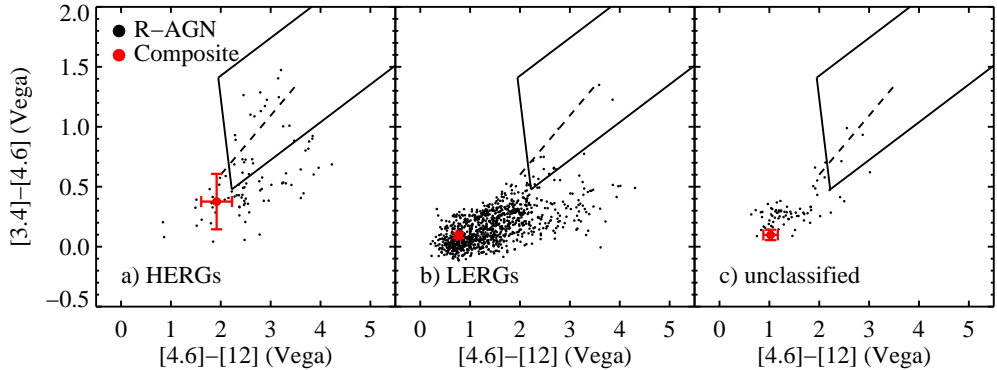}
\caption{WISE color-color plots for the HERGs (left) LERGs (center) and unclassified R-AGN (right). Only R-AGN with a 3-sigma detection in the WISE 12 micron band are shown. The colors from the composite SEDs are also shown. The WISE color selection wedge from \citet{2012MNRAS.426.3271M} is also shown. Samples of AGN selected by their mid-IR colors will miss many radiatively-efficient AGN and be contaminated by radiatively-inefficient AGN. }
\label{colorcolor}
\end{figure*}

\subsection {Mid-IR colors of R-AGN}
Mid-infrared color-color diagrams are often used to identify AGN (Assef et al. 2010, Stern et al. 2012). By examining the WISE mid-IR properties of X-ray selected AGN \citet{2012MNRAS.426.3271M} suggested that a combination of the WISE [3.4]$-$[4.6] and [4.6]$-$[12] colors could be used to identify high-excitation AGN. The WISE [3.4]$-$[4.6] color has been proposed to separate normal galaxies from those with elevated emission from the AGN accretion disk (Assef et al. 2010, Yan et al. 2013), while the WISE [4.6]$-$[12] color can separate early-type galaxies with little warm dust from dusty star formers. \citet{2012MNRAS.426.3271M} found that the WISE 22 $\mu$m band is less effective at selecting luminous AGN than the first three WISE bands because of its shallower depth and because of increased contribution from star-formation activity at the WISE sensitivity.  \citet{2014MNRAS.442..682M} also found an increased contribution from star formation at  24 $\mu$m in a sample of 175 R-AGN in the redshift range $z\leq3.5$.  Since our sample of R-AGN is composed of both high and low-excitation R-AGN, it can be used to evaluate the effectiveness of WISE colors at separating the two modes of accretion. Additionally, the composite SEDs that we generate allow us to investigate the potential biases of AGN studies that use only galaxies that have been individually detected in longer WISE bands, which represent the minority of the BH2012 sample.

Figure~\ref{colorcolor} is a WISE color-color plot of the field R-AGN in the sample, together with the \citet{2012MNRAS.426.3271M} selection wedge. The dashed line shows the WISE blazar strip defined by \citet{2012ApJ...750..138M}. 
We only plot those R-AGN with a 3-sigma detection in the WISE 12 $\mu$m, all of which have 3-sigma detection at 3.4 and 4.6 $\mu$m.
 The WISE colors in these plots have not been dereddened, because the correction is very small. 
The HERGs are shown in panel (a), LERGs in panel (b), and the unclassified R-AGN in panel (c). The detection fraction at 12 $\mu$m for LERGs is 39\%, for HERGs is 81\%, and for unclassified R-AGN is 17\%, while the overall detection fraction is 37\%. 
We have also plotted the results of our Bayesian estimations of the two WISE colors as a red symbol in each panel. The location of the 

In all three plots, it would appear that the composite SED from the Bayesian estimation is significantly bluer than most individually detected R-AGN, as the red point is to the left of what appears by eye to be the average of the individual galaxies. This is due to the fact that we include all galaxies in the Bayesian estimate, which pushes the composite measurement bluer than the observed colors of the higher signal-to-noise detections. 
While some of the HERGs in panel (a) lie within the \citet{2012MNRAS.426.3271M} selection wedge, the majority (65 of the 88 plotted or 74\%) lie outside of it. Since the composite color is outside of the \citet{2012MNRAS.426.3271M} selection wedge the weak and non-detections at 12$\mu$m are probably outside as well. If these are included in counting then the fraction of HERGs outside of the selection wedge increases to 77\%. 

The LERG sample is shown in panel (b), and nearly all of these R-AGN lie outside of the \citet{2012MNRAS.426.3271M} selection wedge in the region occupied by non-AGN, as most early-type galaxies have [3.4]$-$[4.6] colors bluer than 0.5, regardless of whether they host an R-AGN (Yan et al. 2013, Toba et al. 2014). 
This combined with the results from panel (a) suggest that the \citet{2012MNRAS.426.3271M} selection can produce samples of HERGs that are relatively free of LERGs, although such samples will be incomplete. 
Our results support those of \citet{2014MNRAS.438.1149G}, whose analysis of four complete samples of R-AGN suggested that WISE colors cannot be used to select LERGs. This is not surprising, since the lack of an accretion disk and associated obscuring torus in LERGs would preclude strong mid-IR emission. The colors of the unclassified R-AGN in panel (c) are similar to those of the LERGs, suggesting that some of the R-AGN in this category are LERGs. 

\section{Discussion}

\subsection{Feedback, accretion mechanism, and the environment }

We have found that the host galaxies of LERGs have less UV and mid-IR emission than otherwise similar but radio-quiet galaxies, as shown in panel (a) of Fig.~\ref{accretion}. When we subdivide the field LERGs by mass and age, we see that the effect is strongest in the lower mass galaxies (panels (a) and (b) of Fig.~\ref{agemass}). In the lower mass and older age bin the difference between the control and R-AGN SEDs in both FUV and NUV is 0.17 dex (i.e. the SFR of the R-AGN is $\sim$70\% of that of the controls). In contrast for higher mass galaxies the SEDs of LERGs and radio-quiet galaxies are quite similar in the UV and mid-IR portions of the spectrum.
One possible interpretation of this result is that in more massive galaxies it is the so called gravitational quenching \citep{2008MNRAS.383..119D} that dominates, while in less massive it is the radio-mode AGN. Our results are consistent with the picture of LERGs suppressing star formation by preventing gas from cooling onto their hosts from the cosmic web.  

It is quite likely that the majority of massive ellipticals go through cycles of R-AGN activity. 
The duration of R-AGN outbursts may be on the order of 10$^6$ to 10$^7$ yrs (Alexander \& Leahy 1987, Shabala et al. 2008), with the time between outbursts being of about the same duration (Schoenmakers et al. 2000, Konar et al. 2013).
Therefore we do not expect that the galaxies that are currently detected as R-AGN will show the full effect of R-AGN feedback, as in a hypothetical case where ellipticals whose nucleus is never active are compared with those which are always `on'. 
Thus, R-AGN feedback may also suppress SF in more massive galaxies, but the differences between current R-AGN and the control group (currently inactive galaxies) may be erased if the effects of feedback persist in between the active phases, e.g., facilitated by the presence of hot, gaseous halos that are common in more massive galaxies and may store AGN energy \citep{2009ApJS..182..216K}.
We see little or no difference in the UV and mid-IR emission of R-AGN and their controls in denser environments (panels (b) and (c) of Fig.~\ref{enviro}, suggesting that in clusters the gas accretion is already prevented by the hot cluster environment, regardless of the operation of the AGN. Clusters with strong cooling flows may lead to some BCGs having star formation rates that reach tens of solar masses per year, as shown by \citet{2008ApJ...681.1035O}. Our results indicate that such high SFRs due to cooling flows must be uncommon since the UV of the LERG BCGs is comparable to that of the cluster members and field galaxies, as shown in panel (d) of Fig.~\ref{enviro}. 

In contrast to the results for LERGs, we find no evidence of either positive or negative feedback in the hosts of HERGs, as they have similar levels of UV emission to their controls (panel (c) of Fig. 6). 
The HERG hosts have much higher levels of star formation than LERGs (panel (d) of Fig. 6), supporting the idea that they are fueled by large quantities of cold gas that also fuels star formation. 
Our results are consistent with those of \citet{2015A&A...581A..33D}, who found that in a sample of $\sim$80 nearby R-AGN those that had elevated UV emission also had a MIR excess.
In addition to the radio jet, HERGs may also impact their hosts via radiation from their accretion disks. Roos et al. 2015 modeled the impact of AGN ionizing radiation and found that although high-excitation AGN are capable of maintaining their halo gas in an ionized state, the giant molecular clouds that are the sites of star formation are too dense to be affected. Our results for HERGs support this picture, since we find no evidence of feedback in their hosts.

We have shown that the SEDs of HERGs and LERGs are very different in the UV and IR, as shown in panel (d) of Fig.~\ref{accretion}. The HERGs have strongly elevated mid-IR emission from their obscuring tori. They are also more luminous at optical and UV wavelengths. This mirrors the situation at higher redshifts, where low-luminosity R-AGN are hosted by massive ellipticals while more radio-luminous AGN are observed to have large mid-IR and UV excesses (Baldi et al 2014). Our results are also in line with those of Best \& Heckman 2012 who found that for their sample of R-AGN, from which ours is drawn,  HERGs tend to be bluer in SDSS bands than LERGs. We find that this is also true for the GALEX and WISE bands as the HERGs have strongly elevated UV and mid-IR emission compared to LERGs. 
The HERGs also have a strong mid-IR excess relative to their controls, which suggests that the AGN in these galaxies are of the radiatively efficient type. 

\subsection{WISE colors, star formation, and radio luminosity}

\begin{figure}[t!]
  \centering
\includegraphics[width=0.5\textwidth]{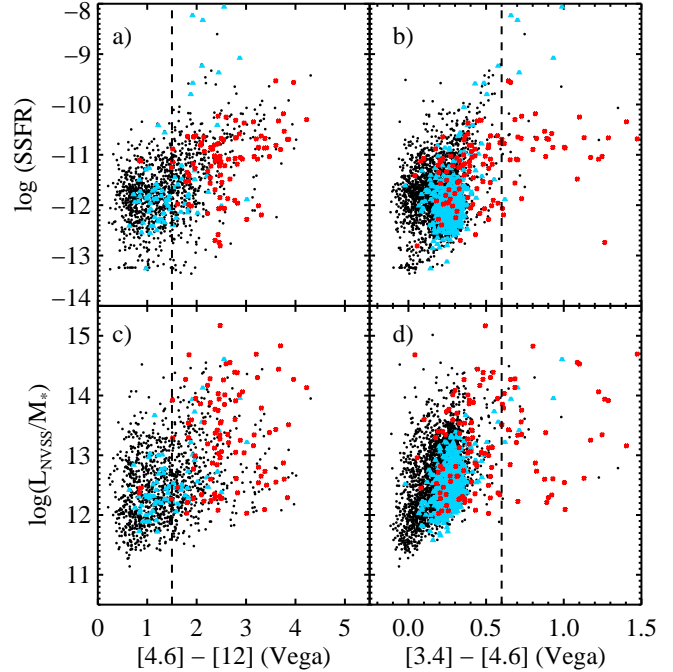}
\caption{WISE [3.4]$-$[4.6] and [4.6]$-$[12] colors plotted against specific radio luminosity and specific star formation rate for LERGs (black circles) HERGs (red stars), and unclassified R-AGN (blue triangles). Only those R-AGN with a 3-sigma detection are shown. The dashed line at [4.6]$-$[12]=1.5 shows the division between spirals and ellipticals, and the line at [3.4]$-$[4.6]=0.6 is the division between powerful AGN and normal galaxies \citep{2010AJ....140.1868W}. These plots show that the unclassified galaxies are broadly similar to the LERGs, and the [4.6]$-$[12] color is more effective at separating HERGs from LERGs.}
\label{wiseradiossfr}
\end{figure}

Our results indicate that samples of HERGs selected by their WISE colors will be both incomplete and contaminated by non-HERGs. \citet{2012ApJ...753...30S} suggested a simple cut at [3.4]$-$[4.6] $\geq$ 0.8 could be used to select reliable and complete samples of AGNs. Of the HERGs shown in panel (a), only 20 of 88 or 23\% are above this line while 3 LERGs and 2 unclassified R-AGN would also be included in such a sample, suggesting that this selection criterion would miss most HERGs. If we use the less stringent color cut at [3.4]$-$[4.6]=0.6 of \citet{2010AJ....140.1868W}, a greater number of HERGs (33 of 88 or 38\% of those plotted) would be selected. However the contamination by non-HERGs would increase, as 18 LERGs and unclassified R-AGN are above the [3.4]$-$[4.6] = 0.6 line. 
The composite HERG [3.4]$-$[4.6] color is at [3.4]$-$[4.6] = 0.4, which further shows that simple [3.4]$-$[4.6] color cuts are not effective at selecting complete and reliable samples of high-excitation AGN.
 This is in line with the results of \citet{2013ApJ...772...26A}, who found that simple color cuts lose reliability at fainter fluxes.

The WISE [4.6]$-$[12] color has been suggested as a way to distinguish between LERGs and HERGs.
\citet{2014MNRAS.438..796S} found that nearly all of the HERGs in their sample had a [4.6]$-$[12] color redder than 2.0. This is also true for our sample, as 82\% of the HERGs are to the right of the [4.6]$-$[12] = 2.0 line. However the composite color in panel (a) falls just to the left of the line, indicating that most HERGs with weak fluxes would be missed by this color cut. Furthermore panels (b) and (c) show that a sample selected with the  [4.6]$-$[12] = 2.0 cut would be composed mostly (73\%) of non-HERGs. 
We conclude that using this color to select HERGs results in high completeness, but very high contamination.

We now consider the more sophisticated two-color selection wedge of \citet{2012MNRAS.426.3271M}. 
In panel (a) of Fig.~\ref{colorcolor} we see that the composite WISE colors of the HERGs in our sample lie outside of the wedge, suggesting that most HERGs will be missed when using this selection technique. The \citet{2012MNRAS.426.3271M} selection wedge is based on the WISE colors of X-ray selected luminous AGN, so there must be some intrinsic difference in the WISE colors of X-ray and radio-selected AGN. This may be due to the anisotropic nature of the mid-IR emission from the obscuring torus (Maiolino \& Rieke 1995, Buchanan et al. 2006). 
Besides being incomplete, samples of HERGs selected via both WISE colors will also be contaminated by LERGs. 
only 23 of the 40 or 58\% of the R-AGN in Fig.~\ref{colorcolor} that fall within the \citet{2012MNRAS.426.3271M} wedge are HERGs, while most of the remainder (14) are LERGs. We conclude that neither simple cuts in [3.4]$-$[4.6] and [4.6]$-$[12] colors nor combinations of these colors are effective at separating HERGs from LERGs. 
Our results are in line with those of \citet{2015MNRAS.449.3191Y}, who compared the WISE [3.4]$-$[4.6] and [4.6]$-$[12] colors of the portion of the \citet{2012MNRAS.421.1569B} sample that is covered by 2MASS and found that HERGs and LERGs occupy different regions of the diagram, with significant overlap between the two populations. While the mid-IR emission of HERGs and LERGs may arise from different processes, the overlap between the two makes it difficult to distinguish between the two types of R-AGN using WISE colors.

To more directly see what drives the mid-IR colors, we plot the SSFR and SRL as a function of the WISE [3.4]$-$[4.6] and [4.6]$-$[12] colors, as shown in Figure~\ref{wiseradiossfr}.
 In this figure the LERGs are shown as black circles, unclassified R-AGN as blue triangles, and HERGs as red stars. For panels (a) and (c) only R-AGN with 3-sigma detections in the WISE 12 $\mu$m band are shown, while in panels (b) and (d) only those with 3-sigma detections in the WISE 4.6 $\mu$m band are shown. The dotted line at [4.6]$-$[12]=1.5 shows the division between spirals and ellipticals, and the dotted line at [3.4]$-$[4.6]=0.6 is the division between powerful AGN and normal galaxies \citep{2010AJ....140.1868W}.

In panel (a) we see a moderate (Spearman's $\rho\sim0.4$) correlation between star formation and the [4.6]$-$[12] color for all AGN. This is similar to the results of \citet{2012ApJ...748...80D}, who matched WISE sources brighter than 1 mJy to the SDSS spectroscopic catalog to investigate the origin of the 12 $\mu$m emission and showed that the WISE [4.6]$-$[12] color can be used to gauge the star formation activity in a galaxy. Panel (b) shows that there is also a moderate ($\rho=0.43$) correlation between SSFR and the [3.4]$-$[4.6] color for HERGs. In panels (c) and (d) we see that the SRL also shows a correlation with both WISE colors, with the stronger ($\rho=0.4$) correlation being to the [3.4]$-$[4.6] color for both LERGs and unclassified R-AGN. 
However, neither the SSFR nor the SRL correlates very tightly with these mid-IR colors. 

In panels (a) and (b) we see that although the HERGs on average may have a higher SSFR than the LERGs, star formation cannot be used to distinguish between the two populations as they both span the range of SSFRs. This is perhaps more clearly seen in panel (b), where a greater number of both HERGs and LERGs are plotted. Both HERGs and LERGs also span the range of SRLs, as shown in panels (c) and (d), indicating that the SRL cannot be used to distinguish between the two types of R-AGN.

\section{Summary}

We have investigated the UV to mid-IR SEDs of an aggregate sample of radio-loud AGN and have compared them to the SEDs of a control sample of galaxies that have the same quality of the data, similar environments and SF histories as the R-AGN sample. We have in addition re-visited the question of R-AGN selection and classification based on mid-IR colors. Our conclusions can be summarized as follows:

\begin{enumerate}
\item Field LERGs in hosts whose mass is below $\log M^*= 11.3$ have less UV and mid-IR emission than similar radio-quiet galaxies, which suggests that LERGs may have a role in suppressing star formation in these galaxies, perhaps as a result of radio-AGN feedback. The suppression of UV is around 0.2 dex and is statistically significant. The potential feedback that we detect may be a lower limit because we see a difference in SFR even though R-AGN probably have a duty cycle, i.e, the ones we now detect as R-AGN are not always on, and vice versa.
In contrast for galaxies whose mass is above $\log M^*= 11.3$ the SEDs of LERGs and radio-quiet galaxies are quite similar in the UV and mid-IR portions of the spectrum, suggesting another suppression mechanism, e.g. the gravitational quenching, or that the effects of AGN quenching persist between AGN duty cycles, possible because of more massive galaxies more often have got gaseous halos.

\item 
The composite SEDs for LERG cluster members and BCGs are very similar to that of their respective control samples, suggesting that R-AGN feedback in denser environments is not necessary to maintain quiescence.

\item The UV emission of R-AGN BCGs is very similar to that of R-AGN in other environments, suggesting that significantly enhanced star formation in BCGs due to cooling flows is uncommon.

\item 
In contrast to the LERGs, we find that the UV emission in HERGs is similar to that of radio-quiet galaxies, suggesting that HERGs have little effect on the star formation in their hosts. The strongly elevated mid-IR emission of HERGs probably originates from the obscuring torus that is present in high-excitation AGN. The differences in UV and mid-IR emission between LERGs and HERGs are also seen when the R-AGN are classified by their specific radio luminosity rather than excitation mechanism.

\item Samples of HERGs selected using either the WISE [3.4]$-$[4.6] color, the [4.6]$-$[12] color, or a combination of both colors will be incomplete and will be contaminated by a significant number of non-HERGs. 
Neither the mid-IR colors nor the SSFR can be used to securely discriminate between LERGs and HERGs.

\end{enumerate}

Our results present strong evidence for suppressed SF in moderately massive non-cluster radio galaxies, which may be the result of AGN feedback. More massive ellipticals and cluster R-AGN show no suppression of SF compared to control groups, which either means that other mechanisms maintain low SF in those galaxies, or that the effect of suppression in such galaxies extends to periods in between the active phases of the central engine, perhaps due to the presence of hot gaseous halos that store the energy released by AGN during the active phases (e.g., Kormendy et al. 2009; Salim et al. 2012). Our characterization of the possible level of radio-mode feedback and identification of the regimes in which it may be important should serve as a benchmark for theoretical studies and simulations.

\section{Acknowledgements}
We thank the anonymous referee for numerous helpful suggestions. This work was partially supported through NASA ADAP award NNX12AE06G.

GALEX (Galaxy Evolution Explorer) is a NASA Small Explorer, launched in April 2003. We gratefully acknowledge NASA's support for construction, operation, and science analysis for the GALEX mission, developed in cooperation with the Centre National d'Etudes Spatiales (CNES) of France and the Korean Ministry of Science and Technology.

This study uses data from the SDSS Archive. Funding for SDSS-III has been provided by the Alfred P. Sloan Foundation, the Participating Institutions, the National Science Foundation, and the U.S. Department of Energy Office of Science. The SDSS-III web site is http://www.sdss3.org/.

SDSS-III is managed by the Astrophysical Research Consortium for the Participating Institutions of the SDSS-III Collaboration including the University of Arizona, the Brazilian Participation Group, Brookhaven National Laboratory, University of Cambridge, Carnegie Mellon University, University of Florida, the French Participation Group, the German Participation Group, Harvard University, the Instituto de Astrofisica de Canarias, the Michigan State/Notre Dame/JINA Participation Group, Johns Hopkins University, Lawrence Berkeley National Laboratory, Max Planck Institute for Astrophysics, Max Planck Institute for Extraterrestrial Physics, New Mexico State University, New York University, Ohio State University, Pennsylvania State University, University of Portsmouth, Princeton University, the Spanish Participation Group, University of Tokyo, University of Utah, Vanderbilt University, University of Virginia, University of Washington, and Yale University.

This publication makes use of data products from the Wide-field Infrared Survey Explorer, which is a joint project of the University of California, Los Angeles, and the Jet Propulsion Laboratory/California Institute of Technology, and NEOWISE, which is a project of the Jet Propulsion Laboratory/California Institute of Technology. WISE and NEOWISE are funded by the National Aeronautics and Space Administration.


\clearpage
\begin{thebibliography}{}
\bibitem[Abazajian et al.(2009)]{2009ApJS..182..543A} Abazajian, K.~N., Adelman-McCarthy, J.~K., Ag{\"u}eros, M.~A., et al.\ 2009, \apjs, 182, 543
\bibitem[Ahn et al.(2014)]{2014ApJS..211...17A} Ahn, C.~P., Alexandroff, R., Allende Prieto, C., et al.\ 2014, \apjs, 211, 17
\bibitem[Alexander \& Leahy(1987)]{1987MNRAS.225....1A} Alexander, P., \& Leahy, J.~P.\ 1987, \mnras, 225, 1 
\bibitem[Allen et al.(2006)]{2006MNRAS.372...21A} Allen, S.~W., Dunn, R.~J.~H., Fabian, A.~C., Taylor, G.~B., \& Reynolds, C.~S.\ 2006, \mnras, 372, 21 
\bibitem[Assef et al.(2010)]{2010ApJ...713..970A} Assef, R.~J., Kochanek, C.~S., Brodwin, M., et al.\ 2010, \apj, 713, 970
\bibitem[Assef et al.(2013)]{2013ApJ...772...26A} Assef, R.~J., Stern, D., Kochanek, C.~S., et al.\ 2013, \apj, 772, 26
\bibitem[Baldwin et al.(1981)]{1981PASP...93....5B} Baldwin, J.~A., Phillips, M.~M., \& Terlevich, R.\ 1981, \pasp, 93, 5 
\bibitem[Barnes \& Hernquist(1996)]{1996ApJ...471..115B} Barnes, J.~E., \& Hernquist, L.\ 1996, \apj, 471, 115
\bibitem[Becker et al.(1995)]{1995ApJ...450..559B} Becker, R.~H., White, R.~L., \& Helfand, D.~J.\ 1995, \apj, 450, 559
\bibitem[Best(2004)]{2004MNRAS.351...70B} Best, P.~N.\ 2004, \mnras, 351, 70 
\bibitem[Best \& Heckman(2012)]{2012MNRAS.421.1569B} Best, P.~N., \& Heckman, T.~M.\ 2012, \mnras, 421, 1569 
\bibitem[Best et al.(2005a)]{2005MNRAS.362....9B} Best, P.~N., Kauffmann, G., Heckman, T.~M., \& Ivezi{\'c}, {\v Z}.\ 2005a, \mnras, 362, 9 
\bibitem[Best et al.(2005b)]{2005MNRAS.362...25B} Best, P.~N., Kauffmann, G., Heckman, T.~M., et al.\ 2005b, \mnras, 362, 25 
\bibitem[Bower et al.(2006)]{2006MNRAS.370..645B} Bower, R.~G., Benson, A.~J., Malbon, R., et al.\ 2006, \mnras, 370, 645 
\bibitem[Bruzual \& Charlot(2003)]{2003MNRAS.344.1000B} Bruzual, G., \& Charlot, S.\ 2003, \mnras, 344, 1000 
\bibitem[Buchanan et al.(2006)]{2006AJ....132..401B} Buchanan, C.~L., Gallimore, J.~F., O'Dea, C.~P., et al.\ 2006, \aj, 132, 401 
\bibitem[Burns(1990)]{1990AJ.....99...14B} Burns, J.~O.\ 1990, \aj, 99, 14
\bibitem[Buttiglione et al.(2010)]{2010A&A...509A...6B} Buttiglione, S., Capetti, A., Celotti, A., et al.\ 2010, \aap, 509, A6
\bibitem[Cattaneo et al.(2009)]{2009Natur.460..213C} Cattaneo, A., Faber, S.~M., Binney, J., et al.\ 2009, \nat, 460, 213 
\bibitem[Charlot \& Fall(2000)]{2000ApJ...539..718C} Charlot, S., \& Fall, S.~M.\ 2000, \apj, 539, 718 
\bibitem[Ciotti et al.(2010)]{2010ApJ...717..708C} Ciotti, L., Ostriker, J.~P., \& Proga, D.\ 2010, \apj, 717, 708 
\bibitem[Condon(1992)]{1992ARA&A..30..575C} Condon, J.~J.\ 1992, \araa, 30, 575 
\bibitem[Condon et al.(1998)]{1998AJ....115.1693C} Condon, J.~J., Cotton, W.~D., Greisen, E.~W., et al.\ 1998, \aj, 115, 1693
\bibitem[Crockett et al.(2012)]{2012MNRAS.421.1603C} Crockett, R.~M., Shabala, S.~S., Kaviraj, S., et al.\ 2012, \mnras, 421, 1603 
\bibitem[Croft et al.(2006)]{2006ApJ...647.1040C} Croft, S., van Breugel, W., de Vries, W., et al.\ 2006, \apj, 647, 1040
\bibitem[Croton et al.(2006)]{2006MNRAS.365...11C} Croton, D.~J., Springel, V., White, S.~D.~M., et al.\ 2006, \mnras, 365, 11 
\bibitem[da Cunha et al.(2008)]{2008MNRAS.388.1595D} da Cunha, E., Charlot, S., \& Elbaz, D.\ 2008, \mnras, 388, 1595 
\bibitem[Dekel \& Birnboim(2008)]{2008MNRAS.383..119D} Dekel, A., \& Birnboim, Y.\ 2008, \mnras, 383, 119 
\bibitem[de Ruiter et al.(2015)]{2015A&A...581A..33D} de Ruiter, H.~R., Parma, P., Fanti, R., \& Fanti, C.\ 2015, \aap, 581, A33 
\bibitem[Donoso et al.(2010)]{2010MNRAS.407.1078D} Donoso, E., Li, C., Kauffmann, G., Best, P.~N., \& Heckman, T.~M.\ 2010, \mnras, 407, 1078
\bibitem[Donoso et al.(2012)]{2012ApJ...748...80D} Donoso, E., Yan, L., Tsai, C., et al.\ 2012, \apj, 748, 80
\bibitem[Dugan et al.(2014)]{2014ApJ...796..113D} Dugan, Z., Bryan, S., Gaibler, V., Silk, J., \& Haas, M.\ 2014, \apj, 796, 113
\bibitem[Dunn \& Fabian(2006)]{2006MNRAS.373..959D} Dunn, R.~J.~H., \& Fabian, A.~C.\ 2006, \mnras, 373, 959 
\bibitem[Dunn \& Fabian(2008)]{2008MNRAS.385..757D} Dunn, R.~J.~H., \& Fabian, A.~C.\ 2008, \mnras, 385, 757 
\bibitem[Edge et al.(2010)]{2010A&A...518L..47E} Edge, A.~C., Oonk, J.~B.~R., Mittal, R., et al.\ 2010, \aap, 518, L47 
\bibitem[Fabian(1994)]{1994ARA&A..32..277F} Fabian, A.~C.\ 1994, \araa, 32, 277
\bibitem[Fabian(2012)]{2012ARA&A..50..455F} Fabian, A.~C.\ 2012, \araa, 50, 455 
\bibitem[Fabian et al.(2003)]{2003MNRAS.344L..43F} Fabian, A.~C., Sanders, J.~S., Allen, S.~W., et al.\ 2003, \mnras, 344, L43 
\bibitem[Fanaroff \& Riley(1974)]{1974MNRAS.167P..31F} Fanaroff, B.~L., \& Riley, J.~M.\ 1974, \mnras, 167, 31P 
\bibitem[Graham(1998)]{1998ApJ...502..245G} Graham, J.~A.\ 1998, \apj, 502, 245 
\bibitem[Granato et al.(2004)]{2004ApJ...600..580G} Granato, G.~L., De Zotti, G., Silva, L., Bressan, A., \& Danese, L.\ 2004, \apj, 600, 580 
\bibitem[G{\"u}rkan et al.(2014)]{2014MNRAS.438.1149G} G{\"u}rkan, G., Hardcastle, M.~J., \& Jarvis, M.~J.\ 2014, \mnras, 438, 1149
\bibitem[Hao et al.(2010)]{2010ApJS..191..254H} Hao, J., McKay, T.~A., Koester, B.~P., et al.\ 2010, \apjs, 191, 254
\bibitem[Hardcastle et al.(2007)]{2007ApJ...662..166H} Hardcastle, M.~J.,  Kraft, R.~P., Worrall, D.~M., et al.\ 2007, \apj, 662, 166
\bibitem[Heckman \& Best(2014)]{2014ARA&A..52..589H} Heckman, T.~M., \& Best, P.~N.\ 2014, \araa, 52, 589
\bibitem[Heckman et al.(1986)]{1986ApJ...311..526H} Heckman, T.~M., Smith, E~P., Baum, S.~A., et al.\ 1986, \apj, 311, 526 
\bibitem[Ho(2008)]{2008ARA&A..46..475H} Ho, L.~C.\ 2008, \araa, 46, 475
\bibitem[Inskip et al.(2008)]{2008MNRAS.386.1797I} Inskip, K.~J., Villar-Mart{\'{\i}}n, M., Tadhunter, C.~N., et al.\ 2008, \mnras, 386, 1797
\bibitem[Karouzos et al.(2014a)]{2014MNRAS.439..861K} Karouzos, M., Jarvis, M.~J., \& Bonfield, D.\ 2014, \mnras, 439, 861
\bibitem[Karouzos et al.(2014b)]{2014ApJ...784..137K} Karouzos, M., Im, M., Trichas, M., et al.\ 2014, \apj, 784, 137
\bibitem[Kauffmann et al.(2008)]{2008MNRAS.384..953K} Kauffmann, G., Heckman, T.~M., \& Best, P.~N.\ 2008, \mnras, 384, 953 
\bibitem[Kauffmann et al.(2003)]{2003MNRAS.346.1055K} Kauffmann, G., Heckman, T.~M., Tremonti, C., et al.\ 2003, \mnras, 346, 1055
\bibitem[Kaviraj et al.(2015)]{2015MNRAS.454.1595K} Kaviraj, S., Shabala, S.~S., Deller, A.~T., \& Middelberg, E.\ 2015, \mnras, 454, 1595 
\bibitem[Konar et al.(2013)]{2013MNRAS.430.2137K} Konar, C., Hardcastle, M.~J., Jamrozy, M., \& Croston, J.~H.\ 2013, \mnras, 430, 2137 
\bibitem[Kormendy et al.(2009)]{2009ApJS..182..216K} Kormendy, J., Fisher, D.~B., Cornell, M.~E., \& Bender, R.\ 2009, \apjs, 182, 216 
\bibitem[Laing et al.(1994)]{1994ASPC...54..201L} Laing, R.~A., Jenkins, C.~R., Wall, J.~V., \& Unger, S.~W.\ 1994, The Physics of Active Galaxies, 54, 201
\bibitem[Lonsdale Persson \& Helou(1987)]{1987ApJ...314..513L} Lonsdale Persson, C.~J., \& Helou, G.\ 1987, \apj, 314, 13 
\bibitem[Magliocchetti et al.(2014)]{2014MNRAS.442..682M} Magliocchetti, M., Lutz, D., Rosario, D., et al.\ 2014, \mnras, 442, 682 
\bibitem[Mainzer et al.(2011)]{2011ApJ...731...53M} Mainzer, A., Bauer, J., Grav, T., et al.\ 2011, \apj, 731, 53
\bibitem[Maiolino \& Rieke(1995)]{1995ApJ...454...95M} Maiolino, R., \& Rieke, G.~H.\ 1995, \apj, 454, 95
\bibitem[Martin et al.(2005)]{2005ApJ...619L...1M} Martin, D.~C., Fanson, J., Schiminovich, D., et al.\ 2005, \apjl, 619, L1
\bibitem[Massaro et al.(2012)]{2012ApJ...750..138M} Massaro, F., D'Abrusco, R., Tosti, G., et al.\ 2012, \apj, 750, 138
\bibitem[Mateos et al.(2012)]{2012MNRAS.426.3271M} Mateos, S., Alonso-Herrero, A., Carrera, F.~J., et al.\ 2012, \mnras, 426, 3271
\bibitem[Mathews \& Brighenti(2003)]{2003ARA&A..41..191M} Mathews, W.~G., \& Brighenti, F.\ 2003, \araa, 41, 191
\bibitem[Meisenheimer et al.(2001)]{2001A&A...372..719M} Meisenheimer, K., Haas, M., M{\"u}ller, S.~A.~H., et al.\ 2001, \aap, 372, 719 
\bibitem[Narayan \& Yi(1994)]{1994ApJ...428L..13N} Narayan, R., \& Yi, I.\ 1994, \apjl, 428, L13
\bibitem[Nulsen et al.(2007)]{2007hvcg.conf..210N} Nulsen, P.~E.~J., Jones, C., Forman, W.~R., et al.\ 2007, Heating versus Cooling in Galaxies and Clusters of Galaxies, 210 
\bibitem[O'Dea et al.(2008)]{2008ApJ...681.1035O} O'Dea, C.~P., Baum, S.~A., Privon, G., et al.\ 2008, \apj, 681, 1035 
\bibitem[Pace \& Salim(2014)]{2014ApJ...785...66P} Pace, C., \& Salim, S.\ 2014, \apj, 785, 66
\bibitem[Peek \& Schiminovich(2013)]{2013ApJ...771...68P} Peek, J.~E.~G., \& Schiminovich, D.\ 2013, \apj, 771, 68
\bibitem[Peterson et al.(2003)]{2003ApJ...590..207P} Peterson, J.~R., Kahn, S.~M., Paerels, F.~B.~S., et al.\ 2003, \apj, 590, 207 
\bibitem[Reviglio \& Helfand(2006)]{2006ApJ...650..717R} Reviglio, P., \& Helfand, D.~J.\ 2006, \apj, 650, 717
\bibitem[Roos et al.(2015)]{2015ApJ...800...19R} Roos, O., Juneau, S., Bournaud, F., \& Gabor, J.~M.\ 2015, \apj, 800, 19
\bibitem[Sadler et al.(2014)]{2014MNRAS.438..796S} Sadler, E.~M., Ekers, R.~D., Mahony, E.~K., Mauch, T., \& Murphy, T.\ 2014, \mnras, 438, 796
\bibitem[Salim et al.(2012)]{2012ApJ...755..105S} Salim, S., Fang, J.~J., Rich, R.~M., Faber, S.~M., \& Thilker, D.~A.\ 2012, \apj, 755, 105
\bibitem[Salim et al.(2014)]{2014ApJ...797..126S} Salim, S., Lee, J.~C., Ly, C., et al.\ 2014, \apj, 797, 126 
\bibitem[Salim et al.(2007)]{2007ApJS..173..267S} Salim, S., Rich, R.~M., Charlot, S., et al.\ 2007, \apjs, 173, 267 
\bibitem[Satyapal et al.(2014)]{2014MNRAS.441.1297S} Satyapal, S., Ellison, S.~L., McAlpine, W., et al.\ 2014, \mnras, 441, 1297
\bibitem[Schoenmakers et al.(2000)]{2000MNRAS.315..371S} Schoenmakers, A.~P., de Bruyn, A.~G., R{\"o}ttgering, H.~J.~A., van der Laan, H., \& Kaiser, C.~R.\ 2000, \mnras, 315, 371 
\bibitem[Shabala et al.(2008)]{2008MNRAS.388..625S} Shabala, S.~S., Ash, S., Alexander, P., \& Riley, J.~M.\ 2008, \mnras, 388, 625 
\bibitem[Shabala et al.(2011)]{2011MNRAS.413.2815S} Shabala, S.~S., Kaviraj, S., \& Silk, J.\ 2011, \mnras, 413, 2815
\bibitem[Silk \& Norman(2009)]{2009ApJ...700..262S} Silk, J., \& Norman, C.\ 2009, \apj, 700, 262 
\bibitem[Singh et al.(2015)]{2015MNRAS.454.1556S} Singh, V., Ishwara-Chandra, C.~H., Sievers, J., et al.\ 2015, \mnras, 454, 1556
\bibitem[Smith et al.(2012)]{2012ApJ...748..123S} Smith, M.~W.~L., Gomez, H.~L., Eales, S.~A., et al.\ 2012, \apj, 748, 123
\bibitem[Stern et al.(2012)]{2012ApJ...753...30S} Stern, D., Assef, R.~J., Benford, D.~J., et al.\ 2012, \apj, 753, 30
\bibitem[Strauss et al.(2002)]{2002AJ....124.1810S} Strauss, M.~A., Weinberg, D.~H., Lupton, R.~H., et al.\ 2002, \aj, 124, 181
\bibitem[Toba et al.(2014)]{2014ApJ...788...45T} Toba, Y., Oyabu, S., Matsuhara, H., et al.\ 2014, \apj, 788, 45
\bibitem[van Breugel \& Dey(1993)]{1993ApJ...414..563V} van Breugel, W.~J.~M., \& Dey, A.\ 1993, \apj, 414, 563
\bibitem[van Breugel et al.(2004)]{2004IAUS..222..485V} van Breugel, W.,  Fragile, C., Croft, S., et al.\ 2004, The Interplay Among Black Holes, Stars and ISM in Galactic Nuclei, 222, 485 
\bibitem[van Velzen et al.(2012)]{2012A&A...544A..18V} van Velzen, S., Falcke, H., Schellart, P., Nierstenh{\"o}fer, N., \& Kampert, K.-H.\ 2012, \aap, 544, A18 
\bibitem[Voit et al.(2015)]{2015ApJ...808L..30V} Voit, G.~M., Bryan, G.~L., O'Shea, B.~W., \& Donahue, M.\ 2015, \apjl, 808, L30
\bibitem[Wright et al.(2010)]{2010AJ....140.1868W} Wright, E.~L., Eisenhardt, P.~R.~M., Mainzer, A.~K., et al.\ 2010, \aj, 140, 1868 
\bibitem[Yan et al.(2013)]{2013AJ....145...55Y} Yan, L., Donoso, E., Tsai, C.-W., et al.\ 2013, \aj, 145, 55
\bibitem[Yang et al.(2015)]{2015MNRAS.449.3191Y} Yang, X.-h., Chen, P.-s., \& Huang, Y.\ 2015, \mnras, 449, 3191 
\bibitem[Zinn et al.(2013)]{2013ApJ...774...66Z} Zinn, P.-C., Middelberg, E., Norris, R.~P., \& Dettmar, R.-J.\ 2013, \apj, 774, 66
\end{thebibliography}
\end{document}